\documentclass{aa}
\input{epsf}
\begin{document}

\def \bt {\beta_{\rm T}}
\def \mp {m_{\rm p}}
\def \msol {{\rm\ M}_\odot} 
\def \kev {\rm keV} 
\def \etal {et al.\ } 
\def \betamodel {\hbox{$\beta$--model} }
\def \kT {{\rm k}T} 
\def \Mg {M_{\rm gas}} 
\def \Mv {M_{\rm V}} 
\def \Rv {R_{\rm V}} 
\def \rc {r_{\rm c}}
\def \xc {x_{\rm c}} 
\def \fg {f_{\rm gas}}

\def \G {{\rm G}}
\def \Lx {L_{\rm X}}

\def \Dzf {\frac{\Del \Omo}{18\pi^{2}\ \Omz}}
\def \Dz {\Delta_{\rm z}}
\def \dA {d_{\rm A}}
\def \Del  {\Delta_{\rm c}(\Omz,\Lambda)}
\def \ho  {{\rm H_{0}}}
\def \MgT {\hbox{$M_{\rm gas}$--$T$} }
\def \MvT {\hbox{$\Mv$--$T$} }
\def \EMT {\hbox{$EM$--$T$} }
\def \LxT {\hbox{$L_{\rm X}$--$T$} }
\def \MvT {\hbox{$\Mv$--$T$} }
\def \RvT {\hbox{$\Rv$--$T$} }
\def \Omo {\rm \Omega_{0}}
\def \Omz {\rm \Omega_{\rm z}}
\def \hc {\rm h_{50}}
\def \EMT {\hbox{$EM$--$T$} }
\def \rxj {\hbox{RX J1120.1+4318 }}
\def \la {\hbox{${\rm \Lambda CDM}$} }

\title{XMM-Newton observation of the distant ($z=0.6$) galaxy cluster
RX~J1120.1+4318} \author{ M. Arnaud\inst{1}, S. Majerowicz\inst{1}, D.
Lumb\inst{2}, D.Neumann\inst{1}, N.Aghanim\inst{3}, A.
Blanchard\inst{4}, M. Boer\inst{5}, D. Burke\inst{6}, C.
Collins\inst{7}, M. Giard\inst{5}, J. Nevalainen\inst{8}, R. C.
Nichol\inst{9}, K. Romer\inst{9}, R. Sadat\inst{5}} \institute{
CEA/DSM/DAPNIA, Service d'Astrophysique, L'Orme des Merisiers B\^at
709, 91191 Gif-sur-Yvette, France \and Science Payloads Technology
Divn., Research and Science Support Dept., European Space Agency,
ESTEC, Keplerlaan 1, Postbus 299, 2200AG Noordwijk, The Netherlands
\and
Institut d'Astrophysique Spatiale, Universit\'e Paris-Sud, F-91405
Orsay Cedex, France \and Laboratoire d'astrophysique de l'Observatoire
Midi-PyrŽnŽes, UMR5572, UPS, 14, Av.  E. Belin, 31400 Toulouse, France
\and Centre d'Etude Spatiale des Rayonnements, 9 avenue du colonel
Roche, BP4346, F-31028 Toulouse, France \and Harvard-Smithsonian
Center for Astrophysics, 60 Garden Street, Cambridge, MA 02138, USA
\and Astrophysics Research Institute, Liverpool John Moores
University, Twelve Quays House, Egerton Wharf, Birkenhead L41 1LD, UK
\and European Space Agency, Research and Scientific Support Division,
ESTEC, Postbus 299 Keplerlaan 1, 2200AG Noordwijk, Netherlands \and Physics Department, Carnegie Mellon University, 5000 Forbes
Avenue, Pittsburgh, PA15213, USA }

\offprints{M. Arnaud, marnaud@discovery.saclay.cea.fr}

\date{Received 3 April 2002}

\titlerunning{XMM-Newton observation of the distant($z=0.6$) galaxy cluster RX
J1120.1+4318} \authorrunning{Arnaud \etal}

\abstract{We report on a 20 ksec XMM observation of the distant
cluster \rxj, discovered at $z=0.6$ in the SHARC survey.  The cluster
has a regular spherical morphology, suggesting it is in a relaxed
state.  The combined fit of the EPIC/MOS\&pn camera gives a cluster
mean temperature of $\kT = 5.3\pm0.5~\kev$ with an iron abundance of
$0.47\pm0.19$.  The temperature profile, measured for the first time
at such a redshift, is consistent with an isothermal atmosphere up to
half the virial radius.  The surface brightness profile, measured
nearly up to the virial radius, is well fitted by a \betamodel, with
$\beta =0.78^{+0.06}_{-0.04}$ and a core radius of $\theta_{\rm c} =
0.44^{+0.06}_{-0.04}~{\rm arcmin}$.  We compared the properties of
\rxj with the properties of nearby clusters for two cosmological
models: an Einstein - de Sitter Universe and a flat low density Universe
with $\Omo=0.3$.  For both models, the scaled emission measure
profile beyond the core, the gas mass fraction and luminosity are consistent
with the expectations of the self-similar model of cluster formation,
although a slightly better agreement is obtained for a low density
Universe.  There is no evidence of a central cooling flow, in spite of
the apparent relaxed state of the cluster.  This is consistent with
its estimated cooling time, larger than the age of the Universe at the
cluster redshift.  The entropy profile shows a flat core with a
central entropy of $\sim 140~\kev~{\rm cm^{2}}$, remarkably similar to
the entropy floor observed in nearby clusters, and a rising profile
beyond typically $0.1$ virial radius.  Implications of our results, in
terms of non-gravitational physics in cluster formation, are
discussed. }

\maketitle

\section{Introduction}

Galaxy clusters occupy a unique position in the scenario
of hierarchical structure formation, as they are still forming today.  The
statistical properties of this evolving population (e.g. the mass
distribution function and the correlations between physical
quantities,  at various redshifts) therefore provide unique constraints
on cosmological scenarios.

In the simplest model of structure formation, purely based on
gravitation, galaxy clusters constitute a homologous family.  Clusters
are self-similar in shape, and predictable scaling laws relate each
physical property to the cluster total mass (or temperature) and
redshift $z$ (e.g. Bryan \& Norman \cite{bryan}; Eke \etal
\cite{eke}).  From observations with ROSAT and ASCA satellites, it is
now well established that this simple model fails to explain all the
observed structural and scaling properties of the nearby cluster
population (Tozzi \& Norman~\cite{tozzi} and reference therein).  The
evolution with redshift of these properties is an essential piece of
information, still largely missing, to reconstruct the physics of the
formation process.  This information is also important for accurate
$\Omo$ estimates based on the evolution of the cluster mass function
(e.g. Oukbir \& Blanchard~\cite{oukbir}).  To compare the results of
flux-limited X-ray surveys with the prediction of the various
theoretical models, it is necessary to understand the relation between
observed quantities, like the cluster temperature and luminosity, and
the virial mass.  Precise $\Omo$ estimate also requires a good
understanding of the survey selection function, which further depends
sensitively on cluster morphology (e.g. Adami \etal~\cite{adami}).

Following the pioneering work of the Einstein medium Sensitivity
Survey (Gioia \etal~\cite{gioia90}), several large, well controlled,
X--ray samples of distant clusters have been assembled in the last
years, using ROSAT observations (see Gioia \cite{gioia} for a review). 
The exceptional sensitivity of XMM, associated with good spectroscopic
and imaging capabilities, now allows the detailed analysis of these
clusters, down to what was only a detection limit with ROSAT. We
present here the XMM observation of RX~J1120.1+4318, a cluster at
$z=0.6$ detected in the bright Serendipitous High-redshift Archival
ROSAT Cluster (SHARC) survey (Romer \etal~\cite{romer}).  This
observation was made in the framework of the XMM-Newton $\Omega$
project, a systematic XMM Guaranteed Time follow-up of the most
distant ($z>0.45$) SHARC clusters (Bartlett \etal~\cite{bartlett}).

The paper is organized as follows.  In Sec.~\ref{sec:analysis}, we
describe the data analysis performed to derive the surface brightness
profile, global temperature and temperature profiles, which are
presented in Sec.~\ref{sec:result}.  In Sec.~\ref{sec:cosmo}, we
compare the physical properties of \rxj with the predictions of the
self-similar model of cluster formation.  The scaled emission measure
and temperature profiles, the gas mass fraction and $\LxT$ relation
are compared to those of nearby clusters.  This study is made for two
cosmological models: an Einstein-de Sitter Universe (EdS, $\Omo=1$)
and a flat low density Universe (\la) with $\Omo=0.3$ and
$\Lambda=0.7$.  In Sec.~\ref{sec:core}, we study the thermodynamical
state of this cluster - possible presence of cooling gas, entropy
content - to further assess the role of non-gravitational processes in
cluster formation.  Section~\ref{sec:conclusion} contains our
conclusions.

Unless otherwise stated all errors on the cluster parameters are at
the $90\%$ confidence level.  A Hubble constant of $\ho = 50~{\rm
km/s/Mpc}$ is assumed.  At the cluster redshift, 1 arcmin corresponds
to $456~{\rm kpc}$ and $561~{\rm kpc}$ for the EdS model and the \la
model, respectively.

\section{Data Analysis}
\label{sec:analysis}

\subsection{Data preparation}

RX J1120.1+4318 was observed for $\sim 20$ ksec on May 8, 2001 with
the EPIC/MOS and pn camera (using the THIN optical blocking filter) in
Full Frame mode (Turner \etal~\cite{turner}; Str\"{u}der
\etal~\cite{struder}).  We generated calibrated event files using the
tasks emchain and epchain of the SAS V5.1.

We discarded the data corresponding to the periods of high background
induced by solar flares (e.g. see Arnaud \etal \cite{arnaud01}).  We
extracted the light curves in the energy band $[10-12]~\kev$ and
$[12-14]~\kev$ for the MOS and pn data respectively.  In these energy
bands, the effective area of XMM is negligible and the emission is
dominated by the particle induced background.  We removed all frames
corresponding to a count rate greater than $15~{\rm ct}/100~{\rm s}$
(MOS data) and $22~{\rm ct}/100~{\rm s}$ (pn data).  After this
selection, the remaining exposure time is $17.6$ ksec, $17.9$ ksec and
$14.2$ ksec for the MOS1, MOS2 and pn observations.

\subsection{Vignetting correction}

The effective area of the XMM mirrors is a function of off-axis angle
and this vignetting effect depends on energy.  An additional
vignetting effect is due to the RGA obscuration for the MOS camera. 
The vignetting calibration data were those available at time of
release of SAS V5.1.

To correct for vignetting effects, we used the method proposed by
Arnaud \etal (\cite{arnaud01}) for spectra, which can be generalized
in a straightforward way to profiles or images (e.g. Majerowicz
\etal~\cite{majerowicz}).  For each event, we computed the
corresponding weight coefficient, defined as the ratio of the
effective area at the photon position and energy to the central
effective area at that energy.  When extracting spectra, image or
surface brightness profile, each event is weighted by this
coefficient\footnote{This vignetting correction can now be done with
task EVIGWEIGHT of SAS V5.2}.  These `corrected' products correspond
to those we would obtain if the detector response were flat.  The
on-axis response can then be used for spectral fitting and to estimate
the physical parameters of the cluster.  We used the spectral response
files {\sf m1$\_$thin1v9q20t5r6$\_$all$\_$15.rsp}, {\sf
m2$\_$thin1v9q20t5r6$\_$all$\_$15.rsp} for EPOC/MOS1,2 and {\sf
epn$\_$ff20$\_$sY9$\_$thin.rmf V6.1} for EPIC/pn
respectively\footnote{see http://xmm.vilspa.esa.es/calibration/}. 
Those files include both the effective area and the redistribution
matrix.

\subsection{Background subtraction}
\label{sec:bkg}

After cleaning for flare events, the XMM background is dominated by
the Cosmic X--ray background (CXB) and the non X--ray background (NXB)
induced by high energy particles.  The former component, which
dominates at low energy, depends on the observed position in the sky and is
vignetted by the optics.  The latter component, which dominates at
high energy, is not vignetted but varies slightly across the detector
and with time.

To subtract the total XMM background, we used EPIC blank sky event
files (one for each camera) obtained by combining several high
galactic latitude pointings.  The data are cleaned for background
flares and bright point sources are excluded (Lumb~\cite{lumb}).  The
sky coordinates in the event files were modified using the aspect
solution of the \rxj observation, so that extraction can be done in sky
coordinates, while insuring that the same detector region is
considered for both the blank field and the source observation.  The
background level, estimated from the total count rate in the whole FOV
in the high energy bands defined above, was found to be 0.85 times
smaller for the source observation than for the blank sky data.  This
is typical of the variations observed since launch (Lumb~\cite{lumb}),
which are explainable by variations in Cosmic Ray shielding as the
magnetosphere is pumped up and down by solar activity.
\begin{figure}[t]
\epsfxsize=\columnwidth  \epsfbox{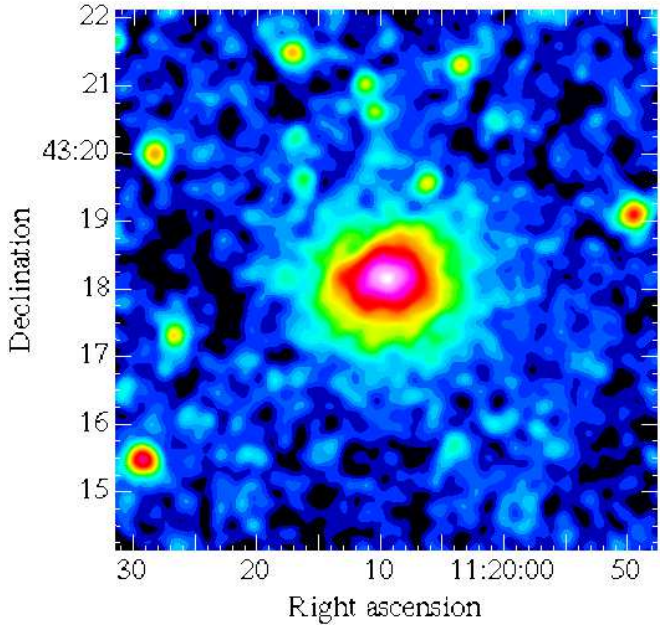} \caption{Combined 
EPIC/MOS1\&MOS2\&pn image of RX J1120 in the $[0.3-5]~\kev$ energy
band (linear intensity).  }
\label{fig:image}
\end{figure}
The background subtraction for each source product (spectrum or
profile) is done in two steps.  The method is fully described in
Appendix~\ref{sec:app}.  We first subtract the corresponding blank
field product, obtained using the same spatial and energy selection,
and normalized by the 0.85 factor defined above.  For consistency, the
blank field products were obtained using the same vignetting
correction method as that used for the source.  The correction factor
is thus formally wrong for the NXB component.  However, as we subtract
blank field and source count rates estimated in the same region in
detector coordinates this correction factor is the same and does not
introduce bias.  This first step thus allows us to subtract properly the
NXB. However, we are left with a residual CXB component, which is the
difference between the CXB in the source region and in the blank
fields (multiplied by the normalization factor above).  This residual
is corrected for vignetting and is expected to be uniform all over the
field of view.  In a second step, we thus subtract this residual
component, using data in the outer part of the FOV, outside the
cluster region.

\section{Results}
\label{sec:result}

The vignetting corrected image in the $[0.3-5] \kev$ energy band is
displayed in Fig.~\ref{fig:image}.  The data of all EPIC cameras are
combined.  The cluster has a regular spherical morphology, suggesting
it is in a relaxed state.

\subsection{Surface brightness profile}
\label{sec:sx}

\begin{figure}[t]
\epsfxsize=\columnwidth \epsfbox{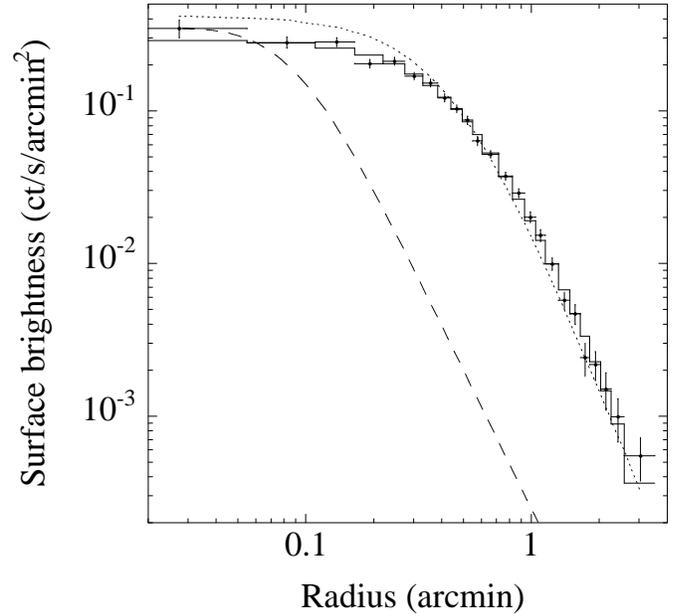} \caption{Combined MOS1,
MOS2 and pn surface brightness profile RX J1120 in the $[0.3-3]~\kev$
energy band.  Dotted line: Best fit \betamodel.  Solid line: same
convolved with the XMM/PSF and binned as the data.  Dashed line: On
axis XMM/PSF, normalized to the central intensity.}
\label{fig:sx}
\end{figure}

We extracted the surface brightness profile of the cluster in the
$[0.3-3]~\kev$ energy band.  This band was chosen to optimize the
signal-to-noise $(S/N)$ ratio.  We cut out serendipitous sources in
the field of view and binned the photons into concentric annuli with a
width of 3.3" (3 pixels of the MOS camera) centered on the maximum of
the X-ray emission for each camera.  Since the cluster is regular,
this center corresponds to the centroid of the emission.  The three
profiles were then summed.  The vignetting correction and background
subtraction was performed as described above.  After subtraction of
the corresponding blank field profile, only a contribution from the
CXB remains (see Appendix~\ref{sec:app}).  Since the data are
corrected for vignetting effects, the CXB surface brightness profile
should be constant with radius.  The profile was indeed found to be
flat beyond $4^\prime$, where we can thus consider that the cluster emission
is negligible.  The residual background was thus estimated from the
data within $4^\prime-7^\prime$ and subtracted from the profile. 
Starting from the central annulus, we re-binned the data in adjacent
annulii so that i) at least a $S/N$ ratio of $3\sigma$ is reached
after background subtraction and ii) the width of the bin increases
with radius, with $\Delta(\theta) > 0.1\theta$.  Such a logarithmic
radial binning insures a $S/N$ ratio in each bin roughly constant in
the outer part of the profile, when the background can still be
neglected.

The resulting surface brightness profile, $S(\theta)$, is shown in
Fig.~\ref{fig:sx}.  The cluster emission is significantly detected up
to $R_{\rm det}=3^\prime$ or $1.37$~Mpc for a critical density
Universe.  Beyond that radius it was not possible to create $S/N>3$
annulus of any width.  The total count rate within the $R<3\arcmin$
region and in the considered energy band is $0.30\pm0.006$ ct/s.

We fitted $S(\theta)$ with a \betamodel convolved with the XMM PSF,
and binned as the observed profile.  The PSF of each camera is modeled
by a normalized King profile, with parameters depending on energy and
off-axis angle (Ghizzardi~\cite{ghizzardi}, Griffiths \&
Saxton~\cite{griffiths02b}).  The overall PSF at each radius is
obtained by summing the PSF of each camera, estimated at an energy of
$1~\kev$, weighted by the respective cluster count rate in the
$[0.3-3]~\kev$ energy band.  The on-axis overall PSF is plotted in
Fig.~\ref{fig:sx} (dashed line).  The convolution with the PSF takes
into account the (small) PSF variation across the region considered. 
However, in practice only the MOS PSF variations are taken into
account.  We used the on-axis pn PSF, due to the lack of available
parametrical fit of the off-axis data.

The fit of the cluster profile gives $\beta =0.78^{+0.06}_{-0.04}$ and
a core radius of $\theta_{\rm c} = 0.44'^{+0.06'}_{-0.04'}$.  The
reduced $\chi^{2}$ is $\sim 1$ ($\chi^{2}=20$ for $21$ d.o.f).  The
convolved best fit model is plotted in Fig.~\ref{fig:sx} (solid line),
together with the corresponding (unbinned and non-convolved)
\betamodel (dotted line).  As can be seen, the PSF mostly affects the
core of the profile.

We then estimated the cluster surface brightness profile corrected for
the effect of the PSF, that we will use in the following
(Sec.~\ref{sec:cosmo}).  In principle, we should have deconvolved the
observed profile.  However since the cluster profile is well fitted by
a \betamodel, we used a simpler procedure.  For each radial bin, we
estimated the ratio of the surface brightness corresponding to the
non-convolved and convolved best fit \betamodel.  The observed surface
brightness profile was then corrected for the effect of the PSF, by
simply multiplying the observed $S(\theta)$ value in each bin by the
model ratio obtained for that bin.

\begin{figure}[t]   
\epsfxsize=\columnwidth \epsfbox{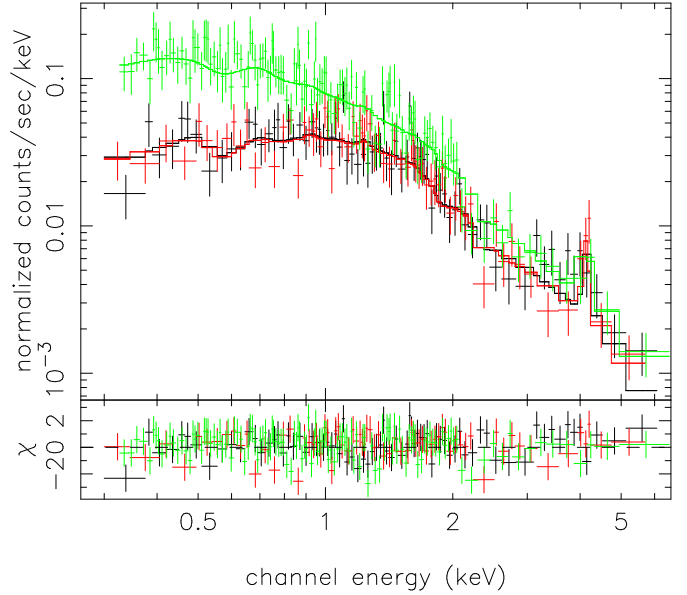} \caption{XMM spectra of
the cluster from the $\theta < 2^\prime$ region.  Black (red) [green]
points: EPIC/MOS1(2)[pn] data.  The EPIC spectra are background
subtracted and corrected for vignetting as described in
Sec.~\ref{sec:analysis}.  Solid lines: best fit isothermal model with
${\rm k}T= 5.3~\kev$, an abundance of 0.47 times the solar value.  }
\label{fig:spec}
\end{figure}

\subsection{Mean temperature}
\label{sec:spec}

The overall MOS1, MOS2 and pn spectra, extracted from the event file,
are shown in Fig.~\ref{fig:spec}.  The spectra are corrected for
vignetting and background subtracted (see Sect.~\ref{sec:bkg}).  To
optimize the $S/N$ ratio, the integration region for the cluster was
restricted to $2^\prime$ from the cluster center, and the residual
CXB spectrum was estimated from the $4^\prime < \theta < 11^\prime$
region.  The spectra are binned so that the $S/N$ ratio is greater
than 3 $\sigma$ in each energy bin after background subtraction.

The spectra are jointly fitted with XSPEC using a redshifted MEKAL
model (Mewe \etal \cite{mewe1},\cite{mewe2}; Kaastra \cite{kaastra};
Liedahl \etal \cite{liedahl}).  We let the relative normalization
between the various instruments be free
but assumed a common temperature and abundance.  When letting the
hydrogen column density $N_{\rm H}$ and redshift $z$ free, we obtained
$N_{\rm H}= 2.2\pm1.2~\times 10^{20}~{\rm cm^{-2}}$, in agreement with
the 21 cm value ($N_{\rm H}= 2.26~\times 10^{20}~{\rm cm^{-2}}$ from
Dickey \& Lockman \cite{dickey}) and $z=0.61\pm0.03$, consistent with
the optical value ($z=0.60$).  We then fixed these parameters to the
21 cm and optical values.  The best fit gives $\kT = 5.3\pm 0.5~\kev$
and an abundance of $0.47 \pm 0.19$.  The reduced $\chi^{2}$ is $\sim
0.94$ ($\chi^{2}=258$ for $276$ d.o.f). 

The temperature values
estimated separately from the MOS and pn spectra are consistent within
the error bars ($\kT = 5.8^{+1.0}_{-0.7}~\kev$ for MOS data and $\kT =
4.5^{+0.8}_{-0.5}~\kev$ for pn data).  We note, however that a lower
temperature is obtained with the pn data than with the MOS data.  This
is likely to be partly due to the remaining calibration systematic
uncertainties in the XMM spectral responses.  Indeed, it is known that
the EPIC-pn and MOS cameras show a relative flux difference of $~4\%$
at low energies which increases with energy above $4.5~\kev$,
resulting in a MOS spectral slope flatter than the pn
(Saxton~\cite{saxton}; Griffiths \etal~\cite{griffiths02a}).
\begin{figure}[t] 
\epsfxsize=\columnwidth \epsfbox{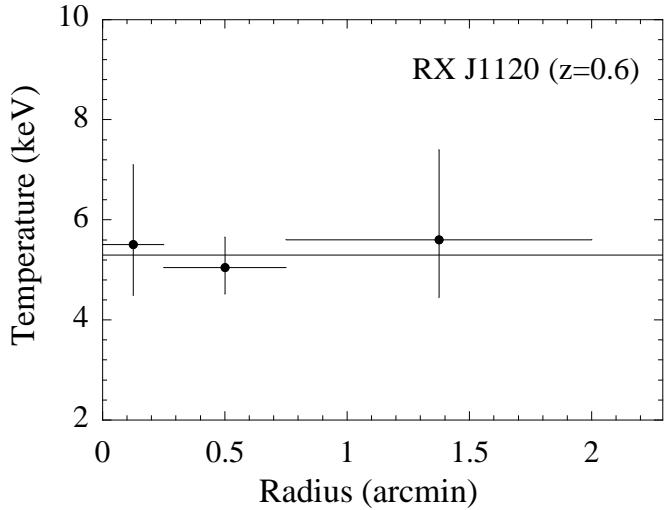} \caption{Radial
temperature profile as a function of angular radius, derived from
XMM/EPIC data.  The horizontal line corresponds to the mean value
derived from fitting the overall spectrum of the region within
$2^\prime$ in radius.  }
\label{fig:ktprof}
\end{figure}

\subsection{Temperature profile}
 
We then extracted the spectra in three concentric annuli, centered on
the cluster X--ray emission peak and fitted the data as described
above.  The corresponding temperature profile is shown in
Fig.~\ref{fig:ktprof}.  It appears flat up to $2^\prime$, within the
error bars.

The emission from the central annulus is affected by the PSF and may
contaminate the other annulii, blurring out gradients if they exist. 
Moreover the energy dependence of the PSF, if not taken into account,
might in principle bias the temperature estimate, since the photon
redistribution is energy dependent.  However, for the MOS1 instrument
for instance, the Encircled Energy Fraction (EEF) within $15"$, the
size of the first bin, is already about $70\%$
(Ghizzardi~\cite{ghizzardi}).  The flux in the first bin is about 1/3
of the second bin flux.  We thus expect a contamination of only $\sim
10\%$ in the second bin and we probably do not significantly
underestimate possible gradients in the central part.  Furthermore,
the EEF varies only by $4.7\%$ between $0.3\kev$ and $5~\kev$, the
minimum and maximum energy for the spectra.  Neglecting the PSF energy
dependence results in a negligible bias ($\sim 0.5\%$) at high
energies.  We emphasize that the XMM telescope PSF has a very weak
energy dependence, so that indeed the correction is tiny and we are
not facing the problem experienced in spectro-imagery with ASCA, for
example.

\section{Scaling Properties}
\label{sec:cosmo}
\begin{figure*}[t]
\epsfxsize=\textwidth \epsfbox{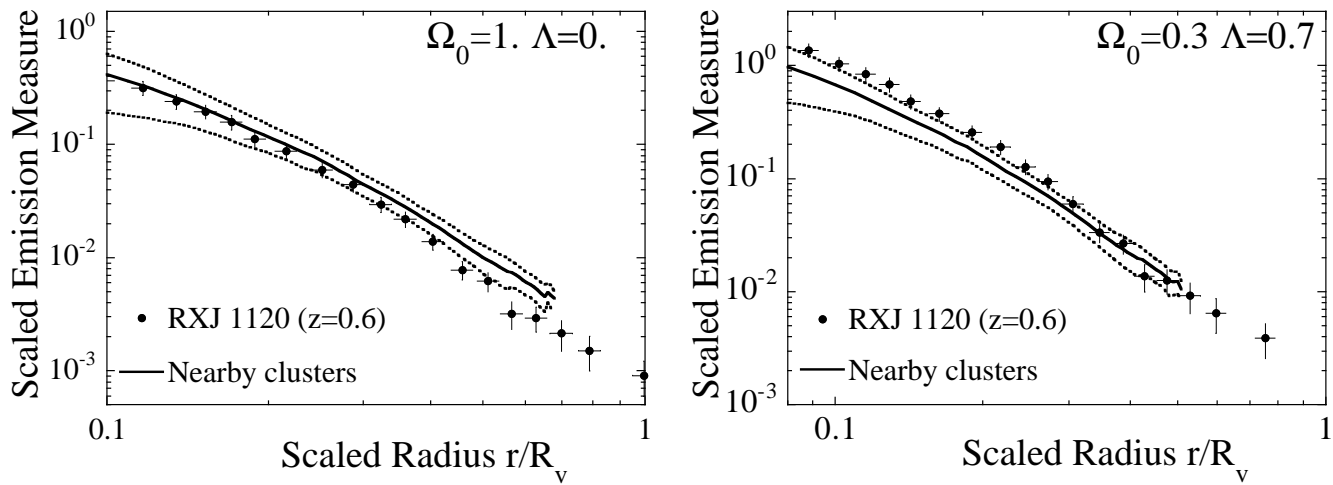} \caption{Comparison
between the scaled emission measure profile of \rxj (data points) and
the mean scaled profile of nearby hot clusters (full line, see text)
for two cosmological models.  The error bars on the scaled variables
take into account the error on the temperature and on the surface
brightness profile.  The dotted lines correspond to the reference
nearby profile, plus or minus the corresponding standard deviation. 
Left panel: Results for a EdS Universe ($\Omo=1, \Lambda=0$).  Right
panel: Results for a \la model ($\Omo=0.3, \Lambda=0.7$).  }
\label{fig:emsc}
\end{figure*}
\subsection{The self-similar model}
\label{sec:selfsimilar}
The self-similar model is based on simple assumptions for cluster
formation, derived from the top-hat spherical collapse model.  The
virialised part of a cluster, present at a given redshift, corresponds
to a fixed density contrast as compared to the critical density of the
Universe at that redshift and the internal shape of clusters of
different masses and $z$ are similar.  Self similarity applies to both
the dark matter component and the hot intra-cluster medium, whose mass
fraction is assumed to be constant.  Consequently simple scaling laws
relate each physical X--ray property, $Q$, to the cluster total mass
(or temperature $T$) and redshift, in the form $Q\propto
A(z)T^{\alpha}$.

This simple model thus makes a definitive prediction in terms of the
evolution of cluster properties.  First, the normalization (but not
the slope) of the $Q$--$T$ scaling relations should evolve with $z$. 
In particular, for a critical density Universe, the virial radius at
fixed $T$ should decrease with $z$ as $(z+1)^{-3/2}$, while the
luminosity and central emission measure increase as $(z+1)^{3/2}$ and
$(z+1)^{9/2}$ respectively.  Furthermore once expressed in scaled
coordinates\footnote{The considered quantity is normalized according
to the scaling relation estimated at the cluster temperature and
redshift and the radius is expressed in units of the virial radius},
the radial profile of any physical quantity (e.g density, temperature)
should be the same at all redshifts.

In this section we will compare the structural and scaling properties
of \rxj to the properties of nearby clusters.  We will consider the
scaled emission measure and temperature profiles, the gas mass
fraction and the $\LxT$ relation.  Results are given for both the EdS
model and the \la model\footnote{The comparison of distant and nearby
cluster properties does not depend on the assumed $\ho$ value: scaled
and physical quantities depend on $\ho$ via a multiplying factor, the
same at all redshifts.}.  The physical properties of \rxj are
summarized In Table~\ref{tab:sum}.

\subsection{The scaled emission measure profile}
\label{sec:emsc}

Recently, Arnaud \etal (\cite{arnaud02}, hereafter AAN) studied the
surface brightness profiles of a sample of 25 distant $(z=0.3-0.83)$
hot $(\kT > 3.5 \kev)$ clusters observed with ROSAT, with published
temperature from ASCA. They found that the scaled profiles of distant
clusters are perfectly consistent with the average scaled profile of
nearby clusters, for a flat low density Universe.  For \rxj, we will
make the same comparison, and we only briefly summarize the method.

\subsubsection{The cluster profile}

We used the surface brightness profile, corrected for the effect of the PSF
(see Sect.~\ref{sec:sx}).  It  is  converted to emission measure
($EM$) profile:
\begin{equation} 
    EM(r) \propto S(\theta)\ (1+z)^{4}/\epsilon(T,z)
    \end{equation}
where $\epsilon(T,z)$ is the emissivity in the energy band considered,
taking into account the instrument response and interstellar
absorption, and $r=\theta\dA$, where $\dA$ is the angular distance. 

The $EM$ profile is then scaled according to the self-similar model. 
We use the standard scaling relations of cluster properties with
redshift and temperature, with the empirical slope of the \MgT
relation derived by Neumann \& Arnaud (\cite{neumann01}).  This
empirical relation is consistent with the observed steepening of the
\LxT relation and reduces significantly the scatter in the scaled
profile of nearby clusters.  The physical radius is thus scaled to the
virial radius, $\Rv$, with $\Rv \propto \Dz^{1/2}\
\left(1+z\right)^{3/2}\ T^{1/2}$ and the emission measures by
$\Dz^{3/2}\ (1+z)^{9/2}\ T^{1.38}$ (see AAN for details).  This
scaling depends on the density parameter $\Omo$ and the cosmological
constant $\Lambda$, via the factor $\Dz =(\Del \Omo)/(18\pi^{2}
\Omz)$, where $\Del$ is the cluster density contrast at redshift $z$
and $\Omz$ the corresponding Universe density parameter.  Analytical
expression of $\Del$ can be found in Bryan \& Norman (1998).  The
scaled profiles also depend on these cosmological parameters via the
angular distance used to convert angular radius to physical radius.

The scaled emission measure profiles of \rxj derived for the EdS model
and the \la model are displayed in the left and right panel of
Fig.~\ref{fig:emsc}, respectively.  The corresponding virial radii are
given in Table~\ref{tab:sum}.  As in AAN, the normalization of the
$\RvT$ relation is taken from the simulation Evrard \etal
(\cite{evrard}).

\subsubsection{Comparison with local data}

For each cosmological model, the scaled profile of \rxj is compared in
Fig.~\ref{fig:emsc} with the corresponding average scaled profile of
nearby clusters.  This profile (full line), and the typical dispersion
around it (dotted lines) is derived, as in AAN, from a sample of 15
hot nearby clusters observed with ROSAT (Neumann \&
Arnaud~\cite{neumann99}).  The comparison is performed at radii beyond
$\sim 0.1 \Rv$, where nearby clusters were found to obey
self-similarity\footnote{A large dispersion was observed in the
cluster core properties, which are dominated by non-gravitational
physics.}.

To be meaningful, this comparison requires that the relative error in
the calibration of XMM/EPIC and ROSAT/PSPC can be neglected.  Our
previous study of A1795 (Arnaud \etal~\cite{arnaud01}), where we
performed a combined fit of the EPIC/MOS and ROSAT/PSPC spectra,
indicates that the fluxes derived by the two instruments match at the
$\sim \pm5\%$ level.  Similar relative calibration uncertainties were
derived by Snowden~(\cite{snowden}).

For both cosmological models, the scaled profile of \rxj is consistent
with the local reference profile, taking into account its intrinsic
scatter.  The \rxj data thus reinforce the validity of the
self-similar model.  However, some significant discrepancy in shape
can be noted.  At large radii the cluster profile falls off more
rapidly than the reference profile.  This corresponds to its higher
$\beta$ value, $\beta=0.78$, as compared to the universal value of
$\beta=2/3$, which fits well the reference curve (Neumann \&
Arnaud~\cite{neumann99}).  The scaled core radius, $\xc= \rc/\Rv$, on
the other hand, is similar to the value ($\xc = 0.12$) of the
reference profile: we obtain $\xc \sim 0.11$ (EdS model) and $\xc \sim
0.14$ (\la model).  This might simply be due to intrinsic scatter in
the properties of distant clusters.  A typical scatter of $\sim 20\%$
is indeed observed in the slopes of nearby cluster profiles (Neumann
\& Arnaud~\cite{neumann99}).  A scatter in cluster properties is
observed as well in numerical simulations of cluster formation
(Navarro, Frenk \& White~\cite{navarro97}).  However, this discrepancy
could also indicate a systematic departures from the self-similar
model considered.  With data on a single cluster, we obviously cannot
distinguish between these two possibilities.

It is also interesting to compare in more detail the results obtained
for the two cosmological models.  The relative position of the scaled
profile of \rxj with respect to the reference profile is higher (in y
axis) for the \la model than for the EdS model.  As discussed by AAN,
the derived scaled profiles of distant clusters depend on the
cosmological parameters, mostly via the angular distance $\dA$.  The
typical dependence is $\propto \dA^{3}$ at fixed scaled radius.  At
$z=0.6$, this $\dA^{3}$ factor is about $85\%$ higher for a \la model
than for an EdS model.  The $\dA^{3}$ dependence is strictly exact for
a profile of logarithmic slope $-3$ (e.g. in the cluster external
region for a \betamodel with $\beta=2/3$).  Due to the steepest slope
of the \rxj profile at large radii, the effect of varying the
cosmological parameters is slightly smaller, of the order of $\sim
60\%$ (Fig.~\ref{fig:emsc}).  The effect is smaller than the intrinsic
dispersion in the nearby profiles at radius below typically $0.2 \Rv$,
and becomes only marginally larger than the $\sim \pm 25\%$ dispersion
at higher radii.  As emphasised by AAN, no definitive conclusion on
the cosmological parameters can then be drawn from the observation of
a single distant cluster.  We simply note that the profile of \rxj is
in better agreement with the local reference profile at large radii
for a \la model than for an EdS model.  In the latter model, the
profile appears to deviate more and more from the reference profile
above $\sim 0.4 \Rv$.  The better agreement observed for the \la model
can simply be an artifact, due to the intrinsic dispersion in the
cluster properties, or again due to systematic departure from the
self-similar model.  However, it is consistent with the statistical
analysis of the scaled profiles of distant clusters, performed by AAN,
which clearly favors a low density Universe.

Finally, we note the statistical quality of the scaled emission
measure profile, as compared to ROSAT/ASCA data.  The profile of \rxj
is traced nearly up to the virial radius, further out than the mean
profile of nearby clusters observed with ROSAT and the errors bars on
most of the data points are smaller than the typical dispersion in the
reference profile.  Note also the very good sampling of the profile. 
Using the same criteria (at least $3\sigma$ detection in each bin),
most of the ROSAT profiles of distant clusters (AAN, Figure 4) are
much more sparsely sampled.

\subsection{The scaled temperature profile}
\begin{figure}[t] 
\epsfxsize=\columnwidth \epsfbox{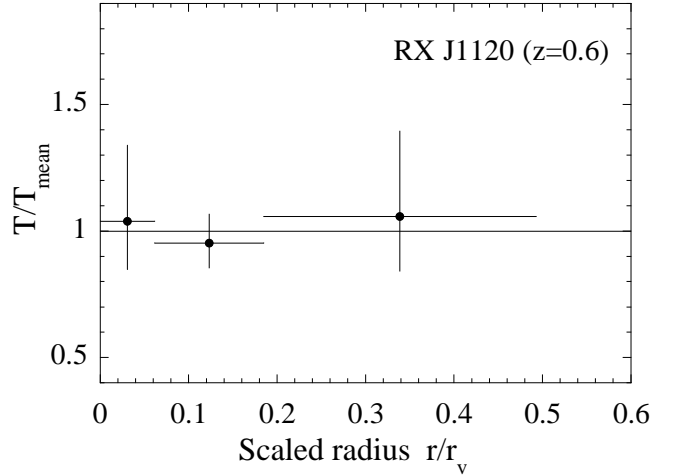} \caption{Scaled
temperature profile.  The temperature is normalized to the mean value
and the radius is expressed in units of the virial radius (\la model,
see Table.\ref{tab:sum}).  }
\label{fig:ktprofsc}
\end{figure}

Cluster temperature profiles can now be measured with high precision
with XMM. First observations of nearby clusters indicate that the
profiles are isothermal (within typically 10\%) up to half the virial
radius (Arnaud \etal \cite{arnaud01}; Arnaud \cite{arnaud01b}).

With XMM we can now also study the evolution of the temperature
profiles.  The cluster temperature profile, normalized to the mean
temperature, is plotted in Fig.~\ref{fig:ktprofsc} as a function of
scaled radius (\la model).  The temperature profile of RX J1120 is
mapped up to about $0.5~\Rv$.  This is the first time that a
temperature profile can be measured at such high redshift.  Although
the errors bars are large, this distant cluster appears isothermal, as
observed for nearby clusters.  This is consistent with the
self-similarity in shape expected in the simple model of cluster
formation.

\subsection{The gas mass fraction}

\subsubsection{VT method}

We first estimated the virial radius ($\Rv$) and mass ($\Mv$) from the
best fit temperature and the theoretical \RvT and \MvT scaling
relations at the cluster redshift, as described in
Sec.~\ref{sec:selfsimilar}.  We recall that these relations correspond
to a fixed density contrast at redshift $z$ and are derived using the
virial theorem (VT):
\begin{equation}
\frac{\G\mu\mp\Mv}{2\Rv}= \bt \kT
\label{equ:VT}
\end{equation}
where the normalization factor is assumed to be $\bt=1.05$, from the
simulation of Evrard \etal (\cite{evrard}).  In that case:
\begin{eqnarray}
\Mv & = & 2.98~10^{15}~\Dz^{-1/2}~(1+z)^{-3/2}
\left(\frac{\kT}{10\ \kev}\right)^{3/2}\msol
\label{eq:mv}
\end{eqnarray}

The best fit temperature corresponds to the emission weighted
temperature within $2'$, which is about half of the virial radius (see
Table~\ref{tab:sum}).  Since $\sim 90\%$ of the emission lies within
that radius, this temperature is a good estimate of the overall
emission weighted value.  Within that radius, there is no significant
evidence of strong gradients.  Nevertheless, if the temperature
profile decreases beyond $2'\sim 0.5~\Rv$, the measured temperature would be
an overestimate of the mass weighted temperature (considered in the
simulations).  However, if the shape of the temperature profile does
not evolve with $z$, the corresponding bias in the VT mass estimate
would be the same at all redshifts and will not affect the following
comparison with nearby cluster properties.

The central hydrogen density, $n_{\rm H,0}$\footnote{$n_{\rm
H,0}=0.85~n_{\rm e,0}$, where $n_{\rm e,0}$ is the central electronic
density for an ionised plasma with the observed abundance} and the
corresponding gas mass within the virial radius, $\Mg$, are derived
from the best fit \betamodel parameters (central surface brightness,
$\beta$ and core radius), assuming a cluster extent equal to the
virial radius.  The errors on these quantities, due to that on the
surface brightness profile, are derived as in Elbaz \etal
(\cite{elbaz}).  The error due to the uncertainty on the temperature,
which appears in the emissivity factor, is negligible.  We emphasize
that the estimate of the gas mass does not require severe
extrapolation of the data and is therefore robust.  The cluster
emission is detected virtually up to the virial radius for an EdS
model ($R_{\rm det} \sim \Rv$) and up to about $0.75~\Rv$ for a \la
model (Table~\ref{tab:sum}).  The gas mass within that radius is
already $70\%$ of the gas mass within $\Rv$.

The central hydrogen density, gas mass, virial radius, virial mass and
gas mass fraction, $\fg = \Mg/\Mv$, are given in Table~\ref{tab:sum}
for the EdS and \la models.  The uncertainty on this last quantity is
dominated by the uncertainty on the temperature, through the estimate
of the virial mass.

\subsubsection{BM method}
We also estimated the total mass using the hydrostatic equilibrium (HE)
equation and the isothermal \betamodel (BM method):
\begin{equation}
    \frac{G~\mu\mp\Mv}{2 \Rv} = 
    \frac{3\beta}{2}~\frac{(\Rv/\rc)^{2}}{1+(\Rv/\rc)^{2}}~\kT
\label{equ:BM}
\end{equation}
where $\beta$ and $\rc$ are the slope and core radius parameters of
the gas distribution (derived Sec.~\ref{sec:sx}).

The corresponding virial mass and radius, corresponding to the same
density contrast as in the VT approach, are given in
Table~\ref{tab:sum}.  Comparing Eq.~\ref{equ:VT} and Eq.~\ref{equ:BM}
and neglecting the terms $(\rc/\Rv)^{2}$, the ratio of the virial
masses (or radius) derived from the VT and BM method is:
\begin{eqnarray}
\frac{\Rv(BM)}{\Rv(VT)} \sim \left(\frac{3~\beta}{2~\bt}\right)^{1/2}~~~~;~~~
\frac{\Mv(BM)}{\Mv(VT)} \sim \left(\frac{3~\beta}{2~\bt}\right)^{3/2}
\end{eqnarray}

The VT and BM methods give very similar results (see
Table~\ref{tab:sum}).  As compared to the VT estimate, the virial
radius and mass, estimated using the BM method, are $\sim5\%$ and
$\sim15\%$ higher, respectively.  The central hydrogen density is
unchanged.  The gas mass within the virial radius is about $\sim5\%$
higher (slightly larger integration region) and the gas mass fraction
is decreased by about $10\%$.

\begin{table}
\caption[]{Summary of the physical properties of \rxj for an EdS model
($\Omo=1$) and a \la model ($\Omo=0.3, \Lambda=0.7$). }
\begin{center}
\begin{tabular}{lll}
\hline
Model& EdS & \la \\
\hline
$\rc$~(kpc) & $203^{+24}_{-21}$ & $250^{+30}_{-26}$ \\
$R_{\rm det}$~(Mpc) & $1.37$ & $1.68$ \\
$n_{\rm H,0}~(10^{-3}~{\rm cm^{-3}})$ &$6.9^{+0.6}_{-0.5}$&$6.2^{+0.5}_{-0.4}$\\
\hline
&\multicolumn{2}{c}{Virial Theorem}\\
\cline{2-3}
$\Rv$~(Mpc) &$1.40\pm0.07$& $2.28\pm0.11$\\
$\Mv~(10^{14}~\msol)$& $5.8\pm0.8$& $9.5\pm1.3$\\
$\Mg(<\Rv)~(10^{14}~\msol)$& $0.85\pm0.04$ & $1.85\pm0.12$\\
$\fg$~(\%) & $14.6\pm2.1$& $19.6\pm3.0$\\
&\multicolumn{2}{c}{HE + isothermal \betamodel}\\
\cline{2-3}
$\Rv$~(Mpc) &$1.47\pm0.08$& $2.40\pm0.12$\\
$\Mv~(10^{14}~\msol)$& $6.1\pm0.9$& $11.0\pm1.7$\\
$\Mg(<\Rv)~(10^{14}~\msol)$& $0.89\pm0.04$ & $1.94\pm0.12$\\
$\fg$~(\%) & $13.3\pm2.1$& $17.6\pm3.0$\\
\hline
$\Lx~(10^{45}~{\rm ergs/s})$& $1.39\pm0.08$& $2.14\pm0.12$\\
$C$ & $1.7^{+0.6}_{-0.4}$ & $2.4^{+0.8}_{-0.5}$ \\
$C_{\rm mod}(z)$& $2.0\pm0.3$ & $1.65\pm0.25$ \\
\hline
\end{tabular}
\end{center}
\label{tab:sum}
\end{table}
\subsubsection{Comparison with the local gas mass fraction}

The gas mass fraction of \rxj can be compared to the $90\%$ confidence
region derived by Arnaud \& Evrard (\cite{arnaud99}, hereafter AE) for
hot ($\kT > 4~\kev$) nearby clusters using the same total mass
estimate methods.  AE found{\footnote{These values were derived for an
EdS model, but the local value is not sensitive to the cosmological
parameters.  The definition of the virial region is also slightly
different but close enough to have no impact on our conclusion: our
definition corresponds to a density contrast of 178 at z=0 for an EdS
model, while the value quoted are estimated at a density contrast of
200}: $\fg = 20.1\mathtt{[-2.5,+3.0]}\%$ using the VT method and $\fg
= 21.5\mathtt{[-3.5,+4.5]}\%$ with the BM method.  For both the VT and
BM methods, the value derived for \rxj is perfectly consistent with
the AE local value for a \la model, but is significantly lower for an
EdS model.

However, the temperature of \rxj ($\sim
5.3~\kev$) is lower than the median temperature of the AE hot cluster
sample ($\sim 8~\kev$).  We are thus not exactly comparing clusters of
same temperature and there is some indication that the gas mass
fraction varies with $T$.  The empirical slope of the \MgT relation
derived by Neumann \& Arnaud~(\cite{neumann01}) corresponds to $\fg
\propto T^{0.44}$, if the classical \MvT relation is not modified. 
This variation is consistent with the variation, $\fg \propto
T^{0.41\pm0.16}$, found by Mohr \etal (\cite{mohr99}) using the virial
mass estimate.  A variation of $\fg$ with $T$ is also readily apparent
in Fig.3 of AE. Correcting the AE $\fg$ values for a possible bias of
$\sim (5.3/8)^{0.44}$, gives a local value of $\fg=16.8\%$ (VT
approach), now in agreement with the \rxj estimate for both a
\la model and an EdS model, although marginally in the latter case.

The gas mass fraction has been proposed from some time
(Pen~\cite{pen}; Rines \etal \cite{rines}) as a novel distance
indicator.  The method is based on the assumed constancy of this
quantity with $z$, while its estimate from X--ray data depends on the
assumed angular distance.  As Mohr \etal (\cite{mohr00}) pointed out,
the region considered to compare $\fg$ at different redshifts must be
fixed in terms of scaled radius, with the evolution of the virial
radius properly taken into account, to avoid biases due to the
variation of the gas mass fraction with physical radius.  The
comparison performed above, where the gas mass fractions are estimated
within the virial radius, does not suffer from this bias\footnote{Note
that the dependence of $\fg$ on $\Omo$ and $\Lambda$ in the present
approach (given in Neumann \& Arnaud~(\cite{neumann00})) is indeed
different than the $\dA^{3/2}$ dependence derived by Pen (\cite{pen})
and Rines \etal (\cite{rines})}.  However, as illustrated above, the
comparison is further complicated by possible variation of $\fg$ with
cluster temperature.  A much better understanding of the $\fg$--$T$
relation, both in the local and distant Universe is required, before
any conclusion can be drawn on the cosmological parameters.

\subsection{The \LxT relation}

\subsubsection{Cluster luminosity}

We computed the bolometric luminosity, $\Lx$, within the virial radius
(given Table~\ref{tab:sum}).  The observed count rate in the [0.3-3]
keV band, obtained by integrating the surface brightness profile up to
the detection radius, is converted to bolometric luminosity using the
best fit MEKAL spectral model and the instrument response.  The
contribution beyond the detection region is estimated from the best
fit \betamodel, but is totally negligible ($0.7\%$ for the \la model). 
The error on $\Lx$ includes both the statistical error on the count
rate and on the temperature.

\subsubsection{Evolution of the \LxT relation} 
As in AAN and in Sadat
\etal (\cite{sadat}), we divided this luminosity by the luminosity
estimated from the local \LxT relation of AE and the cluster
temperature: $\Lx = 1.15~10^{45}(T/6~\kev)^{2.88}$.  The resulting
factor $C$ is given in Table~\ref{tab:sum} for the two cosmological
models considered.  The contribution to the error due to the
uncertainties on $T$ and $\Lx$ are summed quadratically in the log
space.

This $C$ factor can be compared to the evolution of the normalization
of the \LxT relation, $C_{\rm mod}(z) = \Dz^{1/2}\
\left(1+z\right)^{3/2}$, expected in the self-similar model (AAN). 
This theoretical value, estimated at the cluster redshift, is given in
Table~\ref{tab:sum} for the EdS and \la models.  The `error' bars
corresponds to plus or minus the intrinsic scatter in $\Lx$ estimated
by AE. The luminosity of \rxj is in good agreement with the expected
evolution for an EdS Universe, and in the upper range of the expected
evolution for a \la Universe.  For an EdS model, the luminosity is also
marginally consistent with no evolution.

\subsubsection{Comparison with recent works} 

Several groups quantified the evolution of the \LxT relation using
ROSAT/ASCA data (Sadat \etal~\cite{sadat}; Reichart
\etal~\cite{reichart}; Fairley \etal~\cite{fairley}; AAN).  There is a general
consensus that no significant evolution is observed for an EdS model. 
However, as emphasized by AAN, both the luminosity estimate and the
theoretical evolution depend on the assumed cosmology and a different
conclusion is reached for a \la model.  In that case, AAN found a
significant evolution, consistent with the self-similar model.  Our
results agree with this finding.  They are also consistent with the
results of Reichart \etal (\cite{reichart}): they found an evolution of
$C=(1+z)^{(0.91-1.12q_{0})^{+0.54}_{-1.22}}$ in the redshift range
$z<0.5$ or $C=2.0^{+0.6}_{-0.9}$ extrapolated at $z=0.6$ for the \la
model considered here.  This is similar to the value,
$C=2.4^{+0.8}_{-0.5}$, derived for \rxj.

On the other hand, Borgani \etal~(\cite{borgani01b}), using recent
CHANDRA data up to $z=1.26$, did not find any evidence of significant
evolution, for the EdS model but also for the \la model.  We note
first the large uncertainties on the CHANDRA data, especially above
$z=1$.  Moreover, there is a large dispersion, with data points above
and below the local $\LxT$ curve for the \la model.  To illustrate
this point, let us consider the two clusters RX J0848+4456
($z=0.57,\kT=3.6\pm0.5~\kev$) and MS1137.5+6625
($z=0.78,\kT=5.7^{+0.8}_{-0.6}~\kev$).  They are both relaxed
clusters, with precise $\kT$ measurements, and are in the same
redshift and temperature range than \rxj.  The $C$ value of
MS1137.5+6625, $C=2.0^{+0.7}_{-0.8}$, is perfectly consistent with the
expected evolution: $C_{\rm mod}(z)=1.88$ at the cluster redshift and
is similar to the value obtained for \rxj.  On the other hand, the
luminosity of RX J0848+4456 is particularly low for its measured
temperature: $C=0.6^{+0.4}_{-0.2}$, corresponding to a negative
evolution and of course inconsistent with the expected positive
evolution of $C_{\rm mod}(z)=1.6$.  The measured evolution is
sensitive to systematic uncertainties on the temperature ($C\propto
T^{-3}$).  RX J0848+4456 was observed with ACIS-I (Holden
\etal~\cite{holden}), for which there are still large calibration
uncertainties, especially at low energies.  This is a potential worry
for temperature measurements of high $z$ clusters with this
instrument.  The temperature estimate of MS1137.5+6625, observed with
ACIS-S (Borgani \etal~\cite{borgani01b}), is a priori more secure. 
MS1137.5+6625 was also in the AAN sample and the Chandra/ACIS-S data
are consistent with the ASCA data.  First cross calibration studies of
ACIS-S and XMM/EPIC also show a good agreement between these
instruments (Snowden~\cite{snowden}).  Further cross-calibration
studies and accurate measurements on a larger cluster sample are
definitively required to assess if the presently observed dispersion
is real.

\subsection{Consistency of the results}

In our comparisons with the nearby cluster properties, we found that the
scaled $EM$ profile of \rxj, as well as its gas mass fraction and
luminosity, are consistent with the predictions of the self-similar
model of cluster formation.  Since all three observables are not
independent - they are directly related to the gas density - the
consistency of our results is a priori not surprising.  It is, however
not entirely trivial, as we discuss now.

A good match of the scaled $EM$ profile of \rxj with the local
profile, would imply that i) the shape of the gas profile is similar
to that of nearby clusters ii) the $EM$ and thus gas density scales
with $z$ as expected.  Therefore its luminosity (integrated emission
measure profile) and gas mass (integrated density profile) would
naturally be found to follow the standard evolution.  Since the
standard evolution of the gas mass is the same than the evolution
assumed to compute the virial mass, we would also find a gas mass
fraction consistent with the local value.  However, the emission
measure, luminosity and possibly gas mass fraction also depend on the
temperature.  We emphasize that independent studies of the evolution
of these quantities are thus expected to yield consistent results, but
only if one considers a consistent scaling with $T$ for the three
quantities.  The present analysis is consistent in that sense.  As
mentioned above, the theoretical model used to scale the $EM$ profiles
($\Mg \propto T^{1.94}$) is consistent with the slope of the local
$\LxT$ relation (used to normalize the cluster luminosity) and
corresponds to $\fg \propto T^{0.44}$ (used to estimate the local gas
mass fraction at the cluster temperature).

In spite of the general agreement with the self-similar model, some
differences appear when comparing the results of the EdS and
\la models.  The analysis of the gas mass fraction and scaled profiles
would rather favor a \la model.  However for such a model, the
luminosity is in the upper range of the expected evolution and a
better agreement is observed for an EdS model.  This is actually a
direct consequence of the specific shape of \rxj.  The concentration
of the gas distribution of \rxj is slightly larger than the reference
profile, yielding a higher luminosity as compared to the gas mass. 
These differences can also be understood by looking at the scaled
profiles.  As outlined above, the scaled profile of \rxj is in a good
agreement with the nearby profile at high radii for a \la model.  For
an EdS model, the scaled profile lies more and more below the reference
curve at $r \ga 0.4~\Rv$, indicating a lower gas density than expected. 
Since the external regions contribute most to the gas mass ($\sim
75\%$ of the mass lies beyond $0.3\Rv$), this explains why the gas mass
fraction for an EdS model is marginally too low, even as compared to
the corrected AE local value, while the agreement is better for a \la
model.  On the other hand, for an EdS model, the cluster scaled
profile matches better the local profile in the central region (see
Fig.\ref{fig:emsc}), which contributes most to the X--ray luminosity.

\section{Core properties and non-gravitational effects}
\label{sec:core}

\subsection{Radiative cooling}

The XMM data suggest that \rxj does not host a cooling flow.  There is
indeed no indication of a temperature drop in the central bin and a
good fit is obtained with a \betamodel, down to the center, with no
indication of central excess.

We computed the central cooling time, $t_{\rm cool}$ corresponding to
the measured central density: $t_{\rm cool} = 2.9~10^{10}\ {\rm yrs}
\sqrt{T}/n_{\rm H}$, where $T$ is in $\kev$ and $n_{\rm H}$ in ${\rm
cm^{-3}}$ (Sarazin~\cite{sarazin}).  We obtained $t_{\rm cool} \sim
1.~10^{10}\ {\rm yrs} $.  The age of the Universe at the cluster
redshift is $t = 6.5~10^{9}\ {\rm yrs}$ for the EdS model and $t =
1.1~10^{10}\ {\rm yrs}$ for the \la model.  The cooling time, even at
the cluster center where it is close to its minimum value, is thus
similar to or larger than the age of the Universe.  We thus do not
expect a strong cooling flow in the center, in spite of the apparent
relaxed state of the cluster.

\subsection{Entropy and non-gravitational heating}

\begin{figure}[t]
\epsfxsize=8.5cm \epsfbox{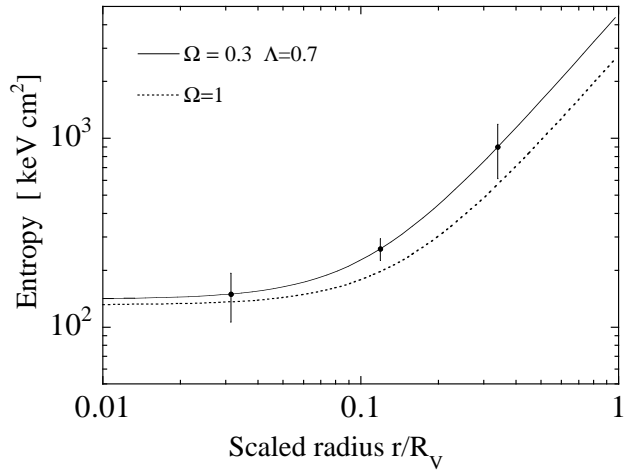} \caption{Entropy profile of
\rxj, derived from the best fit \betamodel and assuming isothermality. 
The radius is scaled to the virial radius for the EdS (dotted line)
and \la (full line) models.  The error bars corresponds to the error
bars on the temperature profile, plotted at the central radii of the
bins considered.  }
\label{fig:entropy}
\end{figure}

As mentioned in the introduction, the simplest self-similar model of
cluster formation fails to explain all the observed properties of the
nearby cluster population.  A definitive evidence of breaking of
self-similarity is the entropy excess (the ``entropy floor'') detected
in the core of cool systems with a baseline entropy of about
$70-140~\kev~{\rm cm^2}$ (Ponman \etal~\cite{ponman}; Lloyd-Davies
\etal~\cite{lloyd}).  The origin of this break of similarity is not yet
understood.

With the present data, we can for the first time study the entropy
profile of a $z=0.6$ cluster, ie look at cluster entropy evolution. 
Note that \rxj appears to be in a relaxed state and the entropy
distribution is not likely to be affected by shocks induced during
recent mergers.  The cluster entropy profile, as a function of scaled
radius, is plotted in Fig.~\ref{fig:entropy} for the EdS and
\la cosmologies.  The entropy $S=T/n_{\rm e}^{2/3}$ is estimated from
the best fit \betamodel and the cluster mean temperature.  The
uncertainty on the profile is dominated by the uncertainty on the
temperature distribution.  Typical errors, corresponding to the
temperature error in each bin of the temperature profile are indicated
in the figure.

One observes a nearly constant entropy level of about $140~\kev~{\rm
cm^2}$ within the central $r<0.1~\Rv$ region and a rising entropy
profile beyond that region.  This shape is not surprising, the entropy
profile shape reflects the shape of the inverse of the density
profile, which has a core radius of about $0.1\Rv$.  The core entropy
is, however, in remarkable agreement with the entropy ``floor'' measured
in nearby clusters.  The same agreement was noted by Arabadjis \etal
(\cite{arabadjis}) in the case of EMSS 1358+6245 at $z=0.33$.

If confirmed on more clusters, this coincidence has potentially
substantial implications for the physics of cluster formation.  We
first note that it is consistent with the expectation of the early
pre-heating scenario, where the gas is pre-heated at a given entropy
level before collapse (e.g. Borgani \etal~\cite{borgani01a}; Tozzi \&
Norman~\cite{tozzi}).  At the beginning of the cluster formation, the
entropy floor prevents shock heating and the collapse is adiabatic. 
The initial entropy is thus preserved in the core.  When the mass of
the system increases, the infall of incoming shells becomes supersonic
and a shock regime begins (gravitational heating).  The external
profile follows the classical 'self-similar' rising profile, as we
observe.  The core entropy of \rxj suggests that the entropy floor was
already established at high $z$: at least $z\ga 0.6$ and probably much
before, since in a hierarchical scenario the core of the cluster must
have collapsed earlier.  Note also that the metal abundance found for
this distant cluster is similar to the abundance observed in nearby
clusters (e.g. de Grandi \& Molendi~\cite{degrandi} and reference therein). 
If early galactic winds are responsible for the gas pre-heating (e.g.
Kaiser \cite{kaiser91}; Evrard \& Henry \cite{evrard91}), we do expect
that this is accompanied by an early enrichment.  Finally, our data
suggest that pre-heating in hot clusters has mostly an effect on the
core properties up to high $z$.  This might explain why AAN verified
self-similarity of the $EM$ profile of hot clusters above $0.1\Rv$ up
to z=0.8.

Radiative cooling has been also proposed to explain the entropy
``floor'': it can remove low entropy gas from near the cluster center,
triggering the inflow of higher entropy material (Pearce \etal
\cite{pearce}).  \rxj, contrary to most nearby relaxed clusters (and
EMSS 1358+6245), is a cluster, for which radiative cooling has
probably not yet affected the core properties (see above).  This
suggests that radiative cooling cannot be the dominant process in the
establishment of the entropy floor.

\section{Conclusion}
\label{sec:conclusion}

XMM-Newton data allow us to measure, with unprecedented accuracy at a
redshift of $z=0.6$, the gas and temperature distribution of the
distant cluster \rxj.  The cluster has a regular spherical morphology,
suggesting it is in a relaxed state.  The cluster temperature is $\kT
= 5.3\pm0.5~\kev$ and there is no significant evidence of temperature
radial gradient up to half the virial radius.  The surface brightness
profile, measured nearly up to the virial radius, is well fitted by a
\betamodel with $\beta =0.78^{+0.06}_{-0.04}$.

The \rxj data reinforce the validity of the self-similar cluster
formation model.  For both an EdS and a \la model, the scaled emission
measure profile beyond $\sim 0.1\Rv$ is consistent with the nearby
cluster reference profile, taking into account its intrinsic scatter. 
Consistently, the gas mass fraction is in agreement with the local
value (although marginally for an EdS Universe), and the luminosity of
the cluster, taking into account its temperature, is consistent with
the expected evolution of the \LxT relation.

There is no evidence of a cooling flow at the cluster center, in spite
of its apparent relaxed state.  This is consistent with the estimated
cooling time, larger than the age of the Universe at the cluster redshift,
indicating that radiative cooling has not yet affected the cluster
properties.  The entropy profile shows a flat core with a central
entropy of $\sim 140~\kev~{\rm cm^{2}}$, remarkably similar to the
entropy floor observed in nearby clusters.  This favors early
pre-heating models for the establishment of this entropy floor.

A statistical study of large sample, as the SHARC cluster sample,
covering a wide range of redshift and luminosity (thus temperature) is
required to better constrain the physics of cluster formation, from
the evolution of cluster properties.  However, the present data
already demonstrate that XMM provides the statistical quality required
by such study.

\acknowledgements

The present work is based on observations obtained with XMM-Newton, an
ESA science mission with instruments and contributions directly funded
by ESA Member States and the USA (NASA).  We acknowledge financial
support from the "Programme National de Cosmologie" funded by INSU,
CEA and CNES. AKR acknowledges partial financial support from the NASA
XMM-Newton GO program.  DJB acknowledges the support for NASA contract
NAS8-39073

 \appendix

\section{Background subtraction method}
\label{sec:app}

In this section we detail the background subtraction method.  This
method applies after screening for flares.  We suppose we have a
`template' event file, for the background estimate, obtained by
collecting several high latitude observations.  In the following, the
data are supposed to be corrected for vignetting effect, as described
in Arnaud \etal (\cite{arnaud01}) and in section 2.2: each event
detected with energy $E$ at location $x,y$ is weighted by a
coefficient $w(x,y,E)$ which is the ratio of the effective area at
position $(x,y)$ to the central effective area, for the energy $E$. 
The Cosmic X-ray Background (CXB) varies across the sky, but can be
considered as uniform at the scale of $30'$, the size of the field of
view.  The non X--ray background (NXB) is not uniform in the FOV, but
is not vignetted.

 For the template file, the corrected count rate, measured at a given
location $(x,y)$ and energy $E$, $T(x,y,E)$, is the sum of the CXB and NXB
contributions:
\begin{equation}
   T(x,y,E) = T_{\rm CXB}(x,y,E) + T_{\rm NXB}(x,y,E) w(x,y,E)
   \label{eq:T}
 \end{equation}
Since the data are corrected for vignetting effects, the CXB component is,
apart from statistical fluctuations, uniform over the FOV and
corresponds to the average CXB for the blank field observations:
\begin{equation}
   T_{\rm CXB}(x,y,E) \equiv T_{\rm CXB}(E)
   \label{eq:TCXB}
 \end{equation}
The NXB component is the NXB for the observations, $T_{\rm
NXB}(x,y,E)$, multiplied by the weight factor.

Similarly for the observation data set, the count rate is:
\begin{eqnarray}
   O_(x,y,E)& =& S(x,y,E)
   + O_{\rm CXB}(x,y,E) \nonumber \\
  &&+ O_{\rm NXB}(x,y,E) w(x,y,E)
   \label{eq:O}
 \end{eqnarray}
 with
\begin{equation}
   O_{\rm CXB}(x,y,E) \equiv O_{\rm CXB}(E)
   \label{eq:OCXB}
 \end{equation}
 where $S(x,y,E)$ is the source contribution and $O_{\rm CXB}(E)$
 corresponds to the CXB at the pointing position of the considered
 observation.  Since the CXB varies across the sky, a priori $O_{\rm
 CXB}(E) \neq T_{\rm CXB}(E)$.  The quiescent NXB can be reasonably
 considered to have the same spectral and spatial characteristics for
 all observations but there is evidence of long term variations
 (D.Lumb~\cite{lumb}).  We can thus write:
\begin{equation}
O_{\rm NXB}(x,y,E) \equiv Q~T_{\rm NXB}(x,y,E)
\label{eq:norm}
\end{equation}
The normalization factor, $Q$, is estimated by considering the total
count rate in the whole FOV in the high energy band, where any cosmic
X--ray emission is negligible.\\

The background subtraction for each source product (spectrum or
profile) is done in two steps.  In a first step, we subtract the
corresponding product of the blank field observation, obtained using
the same energy and spatial selection, normalized by the factor $Q$. 
From Eq.~\ref{eq:T}, Eq.~\ref{eq:TCXB}, Eq.~\ref{eq:O},
Eq.~\ref{eq:OCXB} and Eq.\ref{eq:norm}:
\begin{eqnarray}
  O(x,y,E) - Q~T(x,y,E) &\equiv &S(x,y,E)
 + O_{\rm CXB}(E)  \nonumber \\ && - Q~T_{\rm CXB}(E)
 \end{eqnarray}
and the NXB contribution is thus removed.  The remaining CXB
component, which can be considered uniform, can thus be estimated,
using data outside the source region, where:
\begin{eqnarray}
  O(x,y,E) - Q~T(x,y,E) &\equiv &O_{\rm CXB}(E) - Q~T_{\rm CXB}(E)
 \end{eqnarray}
It can then subtracted in a second step.\\

To illustrate the method, let us consider that we want to estimate the
source spectrum in a particular region, $Reg$:
\begin{equation}
    S_{\rm Reg} (E) = \sum_{\rm Reg}S(x,y,E)
 \label{eq:source}
 \end{equation}
In a first step, we extract the region spectrum from the observation
data set.  The spectrum extracted in the same region from the template
file is then subtracted, after normalization by the factor $A_{\rm
norm}$ above.  The resulting spectrum is:
\begin{eqnarray}
I_{\rm Reg}(E)& =& \sum_{\rm Reg} O(x,y,E) - Q \sum_{\rm Reg} T(x,y,E)\\
 &\equiv& S_{\rm Reg} (E) \nonumber \\
&& + A_{Reg}
(O_{\rm CXB}(E) - Q~T_{\rm CXB}(E))
\label{eq:step1}
 \end{eqnarray}
where $A_{Reg}$ is the surface of the extraction region.  We are thus
left with a residual X-ray background component, which is the estimate
of the difference between the CXB spectrum in the observation and in
the template file (multiplied by the normalization factor).

In a second step, we do the same for a region $Reg'$ outside the
source.  The resulting spectrum is:
\begin{equation}
I_{\rm Reg'}(E) \equiv A_{\rm Reg'} (O_{\rm CXB}(E) - Q~T_{\rm CXB}(E))
\label{eq:step2}
 \end{equation}
The source spectrum can thus be estimated by subtracting this spectrum
to the spectrum obtained in Step 1, after normalization to the size of
the extraction region.  From Eq.~\ref{eq:step1} and Eq.~\ref{eq:step2}
one get:
\begin{equation}
I_{\rm Reg}(E) - \frac{A_{\rm Reg}}{A_{\rm Reg'}} I_{\rm Reg'}(E) \equiv
S_{\rm Reg} (E)
\end{equation}

The same method can be applied to extract the source surface
brightness profile, in a given energy band.  We extract the surface
brightness profile from the observation data set.  The surface
brightness profile extracted in the same energy band from the template
file is then subtracted, after normalization by the factor $Q$ above. 
The resulting surface brightness profile is an estimate of:
\begin{equation}
    I (r) \equiv S (r) + (S_{O_{\rm CXB}} - Q~S_{T_{\rm
    CXB}})
 \end{equation}
 where $S(r)$ is the source surface brightness profile and $S_{O_{\rm CXB}}$
 and $S_{T_{\rm CXB}}$ are the CXB surface brightness in the energy band
 considered for the observation and template file respectively.  The
 residual CXB contribution can be estimated from the measured profile,
 $I (r)$, in the region outside the source and subtracted.

{}

\end{document}